\documentclass{llncs}
\usepackage{geometry}
\usepackage{makeidx}  
\usepackage{amsmath}
\usepackage{amsfonts}
\usepackage{amssymb}
\usepackage{graphicx}
\usepackage{array}
\usepackage{multicol}
\usepackage{caption}
\usepackage{epstopdf}
\usepackage{listings}
\usepackage{hyperref}

\begin{document}
\frontmatter          
\pagestyle{headings}  
%

\vspace{1cm}
%
%
%
%
%
%
\mainmatter              
\title{Considerations about multistep community detection}
\titlerunning{Multistep community detection}  
%
\author{ Cristian Bisconti\inst{1} \and Angelo Corallo\inst{1}
 \and Laura Fortunato\inst{1} \and \underline{Antonio A. Gentile}\inst{1}}
\authorrunning{Antonio Gentile et al.} 
%
%
\institute{Dept. of Innovation Engineering, University of Salento, 73100 Lecce, Italy,\\
\email{antonio.gentile@unisalento.it},\\ WWW home page:
\texttt{http://emi.unisalento.it/sna} 
}

\maketitle              
\begin{abstract}
The problem and implications of community detection in networks have raised a huge attention, for its important applications in both natural and social sciences. A number of algorithms has been developed to solve this problem, addressing either speed optimization or the quality of the partitions calculated. In this paper we propose a multi-step procedure bridging the fastest, but less accurate algorithms (coarse clustering), with the slowest, most effective ones (refinement). By adopting heuristic ranking of the nodes, and classifying a fraction of them as `critical', a refinement step can be restricted to this subset of the network, thus saving computational time. Preliminary numerical results are discussed, showing improvement of the final partition.  
\end{abstract}
\section{Introduction}
\label{intro}
The framework of network analysis has proven a powerful tool in the study of complex phenomena, with applications ranging from biological and social systems, up to technological ones \cite{net.brandes05}. Gaining insight on the structure and behaviour of the network may often be considered a sort of \textit{data fusion} problem, whenever huge data-sets are available. \\
Among the various strategies, outlined to understand large-scale structures, a successful one has pointed out the natural tendency of real-world networks to form \textit{clusters}\footnote{In the literature, for this same concept, also the following terms are equivalently used: \textit{communities, groups, modules, partitions}. For slight preferences in the usage of these terms, refer to \cite{fortu10}}: 
groups of nodes densely connected among them, with sparser links to the rest of the network. Even though the concept is intuitively clear, an operational definition of a `network cluster' is itself under debate: for a concise review of suggested definitions, see \cite{info.fortu04}. Identifying these dense structures inside a network may be crucial for a wide variety of reasons \cite{intro.newman11}\cite{bio.jeong00}\cite{web.broder97}. In other cases, it is already important the mere evaluation of the tendency to form clusters, without detecting cluster members. In fact, this tendency has been found indicative of robustness and stability of the network \cite{info.rosvall07}.
 
The importance of these applications has led recently to the intense development of algorithms, aiming to solve automatically the detection of communities, or to check for the clusterability of the network \cite{fortu10}. The focus is here on the specific case of \textit{community detection}, where number and size of the clusters are free parameters of the problem \cite{mod.newman06}, which addresses also the issue of determining if a \textit{good} partitioning is achievable.  \\
On a different basis, one could distinguish among classes of algorithms, grouped according to their focus. A first class, devoted to capturing the global picture of the network clustering, aiming at a fast solution of the clustering problem given, which especially suits large networks. Such algorithms will be generically indicated in the following as \textit{coarse grain}, since in general they use global metrics as the figure of merit to optimize\footnote{E.g. the \textit{optimization methods} using: \textit{E/I ratios} \cite{eiratio.krack88}, information-compression measures \cite{info.rosvall07}, ..., Hamiltonian-like quantities (spin-hamiltonians \cite{multires.ronhovde09}, \textit{modularity} \cite{mod.newman06}, ...). Another good example is the class of methods known as \textit{block modeling} \cite{fortu10}.}, and often embed approximated methods \cite{mod.cnm04} \cite{fast.blondel08}, thus potentially leading to a relatively high rate of misclassified nodes (e.g. see \cite{critiq.good10}). 
On the opposite side, fine grain algorithms, in particular those involving metrics at the node/edge level\footnote{Like the \textit{edge betweenness}, \textit{information centrality}, other cost functions, directly referred to the network structure (Kernighan-Lin approach \cite{ker.lin70}, ...), or real-world analogies: \textit{current-flow}, \textit{message-passing}, ... \cite{intro.newman11}}, or \textit{hierarchical} structures: in this case, the aim is a precise assignment of the single nodes to the various communities. Moreover, these \textit{refinement} algorithms frequently adopt `exact' methods, for the optimization task they deploy.   

The purpose of this paper is to provide a strategy, enabling to bridge these two different classes. At the moment, in fact, the norm is the straightforward application of a single step algorithm \cite{fortu10}, or multi-step approaches with different optimization schedules for the same metric (see \cite{eigen.newman06} for an example with \textit{modularity}). There is a reason behind this tendency. Small networks can efficiently rely on time consuming algorithms, thus making superfluous to adopt faster methods. These last ones are instead the only feasible chance for large networks. In this paper, we envisage that it is possible to overcome this difficulty, by running a refinement step on only a fraction of the whole network. This fraction is identified via \textit{heuristic metrics}: we call them `heuristic' because, as better shown in the following, the metric chosen not only draws on the characteristics of the network analyzed, but must rely on some `preliminary' clustering results, as computed via coarse algorithms.  \\
In Par.\ref{framework}, after a brief introduction on the framework of our proposal, we will provide a detailed assessment of general features and applicability of our multi-step scheme, and discuss a few metrics which may be adopted as heuristics. Characteristics and a first testing of the method, based on heuristics proposed, will be illustrated in Par.\ref{test}. Some remarks and outlines of future developments conclude this work.
\section{Framework and Methods}
\label{framework}
In the following, we are going to use concepts and metrics derived from graph theory, assuming that: 
\begin{proposition}
The network to analyze can be represented by a graph $\mathcal{G}$.
\label{propg}
\end{proposition}
For $\mathcal{G}$ we adopt the following synthetic definition (see \cite{graphintro.west} for more details):
\begin{definition}
A (directed) (weighted) graph is the ordered pair $\mathcal{G}(V, E)$, with $V$ and $E$ respectively the $n$ vertices ($v_i$) and $m$ edges belonging to $G$. If (directed), the edges $\{v_i,v_j\}$ are ordered pairs. The (weighted) values of the edges among vertices can be embedded in an `adjacency matrix' $A_{ij}$ of dimension $n$, where: $a_{ij}=0$ iff there is no edge linking $v_i$ to $v_j$.
\label{def.graph}
\end{definition}
$\mathcal{G}$ may be a \textit{digraph}\footnote{For a detailed review of peculiarities of clustering approaches in directed graphs, the interested reader may refer to \cite{digraph.malliaros13}.}. This case is explicitly analyzed in the following, where the ordering in the indexes of matrix $A$ (in this case non-symmetric) will be supposed to follow the rule: edge $i\rightarrow j$ is embedded in the element $a_{ij}$, and viceversa. Also the case of a \textit{multimodal graph} can be treated in principle, supposing that a clustering problem, as in Def.\ref{propc}, is well posed for the graph considered.

\begin{figure}
	\centering
		\includegraphics[width=13cm, height=2.5cm]{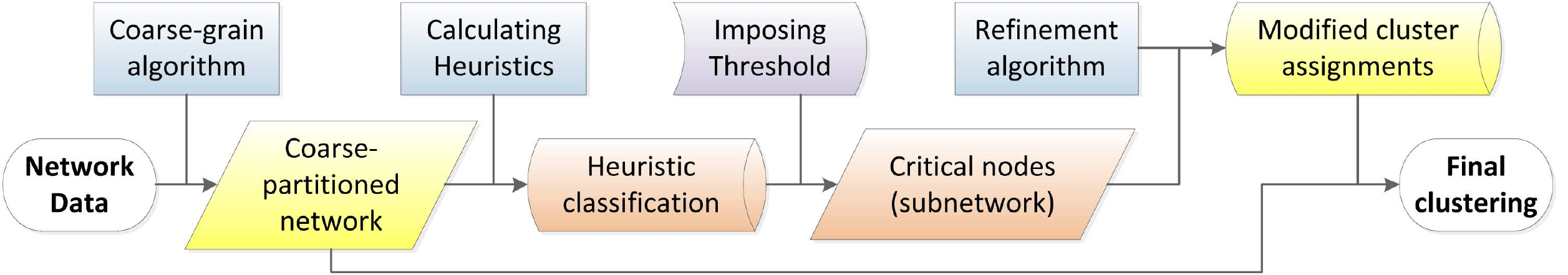}
	\caption{Flowchart of the multi-step method proposed in the text. In blue/violet are distinguished the operational steps. Other colors are referred to generic data (with database symbols) and data in graph-structure.}
	\label{fig:scheme}
\end{figure}

As pointed out in the introduction, in this paper we intend to address the general problem of  automatic clustering in a network. Therefore, we will mainly refer to the case of:
\begin{definition}
`Community detection' as the optimal partition problem\footnote{We stress not to confuse it with the \textit{graph partitioning}, i.e. a specific clustering problem, see \cite{fortu10}} of finding $\mathcal{C}_k$ non-overlapping, non-empty components (i.e. subgraphs) of $\mathcal{G}$: $\bigcup_k \mathcal{C}_k=\mathcal{G}$. Their number $k$ may be an input of the problem, or left as a free parameter. 
\label{propc}
\end{definition}
Prop.\ref{propg} and Def.\ref{propc} are strictly required, for the following discussion to make sense. The assumption of `no-overlap', instead, may be (partially) relaxed, though leading to interesting applications. Moreover, if the partitions $\mathcal{C}_k$ optimize a `quality function' \footnote{Which may equivalently be a `cost function': a survey of such functions is available in \cite{fortu10}.}, it would be eased a quantitative comparison among different solutions (eventually found at different steps, or by different combinations, of the global scheme as in Fig.\ref{fig:scheme}). Further comments about these assumptions can be found in \cite{mmbdft14}.
\subsection{The multi-step scheme}
\label{multistep}

Our contribution for a multi-step approach is the proposal of \textit{heuristic metrics}, that have both low computational time-complexity, and a good efficiency in classifying the nodes according to their degree of membership to possible communities. These metrics enable the adoption of a scheme including: 
\begin{enumerate}
	\item a \textsl{coarse grain algorithm} for the initial clustering guess,
	\item an efficient, \textsl{heuristic metric} for the retrieval of a reduced set of nodes, requiring further analysis upon cluster assignment,
	\item a \textsl{refinement algorithm}, to be run on the nodes produced by the previous step, i.e. a fraction of the initial graph, to improve the `quality' of the final partitions. 
\end{enumerate}
These three main elements, along with some other features which will be introduced in the text, are in Fig.\ref{fig:scheme}, which shows the global multi-step structure. 

The very same introduction of a \textit{refinement} brings along the problem of identifying a measure, able to compare different clustering solutions (i.e. a \textit{relative} measure). A general discussion about the problem is clearly outside our scope: additional information can be found in \cite{clust.kann04},\cite{clust.qual.kri},\cite{part.qual.rob}. \\
In the following, the ultimate target will be to approximate those partitions, which would be provided if the fine-grain analysis chosen was to be performed on the whole network (implicitly assumed to be the best partitioning available). About the distance among partitions provided at different steps in our procedure, we adopt the ratio between the minimum number of elements to delete from a graph $D(P, P')$, so that the two induced partitions become identical, and the size of the graph \cite{mmbdft14}:
\begin{equation} 
 D(P, P'):= n_D(P, P')/n
\label{part.dist}  
\end{equation}
A multi-step procedure requires a few qualitative hypotheses, in order to be effective:
\begin{proposition} refinement algorithm used must be able to perform displacements of single nodes;
\label{assmp}
\end{proposition}
\begin{proposition} 
the heuristic metric chosen must perform as a good figure of merit, in quantifying the `criticality' of the nodes in the network;
\label{assmp2}
\end{proposition}
\begin{proposition} 
however chosen, the fastest method (eventually approximate) to compute the metric in Prop.\ref{assmp2} should at least outperform, in time-complexity, the refinement clustering algorithm.
\label{assmp3}
\end{proposition}
Prop.\ref{assmp} derives from the necessity, once a node-ranking has been established, to analyze, and eventually modify, the cluster attribution of specific nodes, so to address the way the refinement algorithm works. The second statement emphasizes how a perfectly efficient metric should rank first \textit{only} those nodes which will be misassigned by the coarse grain algorithm. Clearly, given that a variety of algorithms could be used as coarse-grain, this `perfect efficiency' is indeed a relative concept, and independently from the refinement algorithm there is no way to define it. Finally, for an alternative clustering procedure to be competitive, with respect to the refinement algorithm, all of its steps must be (much) faster to compute, as stated in Prop.\ref{assmp3}.  

A first naive approach, for retrieving the critical nodes of the network, could be to adopt \textit{centrality} measures \cite{sna.wasser94} from network theory. However, there are a few drawbacks \cite{mmbdft14}, such as the implicit assumption, that the refinement should involve the most `important' nodes. Misclassifying a central node is likely more problematic, but there is no general reason why the coarse-grain algorithm should perform worse on most central nodes. \\
Let us introduce a few qualitative statements, aiming to satisfy the requirements in Propp.\ref{assmp}-\ref{assmp3}. As the first, the assignment of a node to a cluster depends the distribution of its links to neighbour nodes \cite{fortu10}: this leads to introducing the node degrees. In order not to relate the heuristic to the importance of the node, some normalization factor must be introduced. In undirected graphs, we will use a total `symmetrized'  degree for each node $j$, $d_T(j):=\sum_{i}{\left( {a}_{ij}+{a}_{ji} \right)}/2$. \\ 
Given the hypothesis of computing heuristics only after a first coarse assignment of nodes to clusters, one is able to distinguish among edges \textit{inside} or \textit{outside} a given cluster, via the binary function $com$ with values in $\{-1,1\}$:
\begin{equation} com(i,j):=
	\begin{cases} 
	 -1 \;\;\;\;\;\; & \mbox{(if $i$ and $j$ belong to different communities )} \\
	 +1 \;\;\;\;\;\; & \mbox{(if $i=j$ $\vee$ if $i$ and $j$ belong to same cluster)}
	\end{cases}
	\label{comd}
\end{equation}
We claim that a 1$^{st}$ order heuristic metric, suitable for quantifying the criticality of node $j$, can be formulated as:
\begin{equation}
 H_1(j) = \frac{1}{2d_T(j)} \sum_{{i}} \left( {a}_{ij}+{a}_{ji} \right) {com(i,j)} \\
\label{heu1}
\end{equation}
while for the 2$^{nd}$ order heuristic we suggest:
\begin{equation}
 H_2(j) = \frac{1}{2 d_T^2(j)} \sum_{i \neq j} \left(a_{ij}+a_{ji} \right) \;  com(i,j) Q \; d_T(i) H_1(i) \\
\label{heu2}
\end{equation}
where:
\begin{equation} 
 Q=\frac{\delta(\mathcal{G})}{\Delta(\mathcal{G})} 
 \label{qd}
\end{equation}
is a normalization factor, with $\delta(\mathcal{G})$ and $\Delta(\mathcal{G})$ the minimum and maximum degree of the nodes in $\mathcal{G}$, respectively.\\
A few remarks. The expressions about the \textit{order} refer to the width of the network sample taken into account for each node: the edges shared \textit{with} its neighbour nodes in the 1$^{st}$ case, and also all edges shared \textit{by} its neighbour nodes in the 2$^{nd}$.\\
Both heuristics are bounded: as it is easy to verify, $-1 \leq H_1, \; H_2 \leq +1$. Thus, the first order heuristic may be interpreted as a normalized measure of the \textit{correlation} of the node with its cluster of assignment, disregarding its neighbour nodes. Evidently, a positive correlation is here an index of robust assignment, whereas negative correlations indicate misassignment. \\
Qualitatively, re-introducing in (\ref{heu2}) the heuristic $H_1$ accounts for the cluster assignment of neighbour nodes: the stronger the connection of a neighbour node $i$ to its own cluster, the higher we expect its contribution to the (mis)assignment score of analyzed node $j$, if $com(i,j)=+1$ $(-1)$. The factor
\begin{equation}
 M := Q d_T(i) / d_T(j) ,
	\label{terms2}
\end{equation}
instead, can be interpreted as a measure relating the contribution from node $i$ to its relative `importance' in the network, compared to node $j$ (thus the presence of $Q$). That is, $M$ reduces the contribution from $H_2$, compared to $H_1$: if $d_T(i) / d_T(j) = \rho \Rightarrow M < \rho^2$. 

Another interesting point to analyze is how to combine the two heuristics introduced. We suggest that the most profitable figure of merit is the convex combination:
\begin{equation}
  H(j):= \alpha H_1(j)+ (2-\alpha) H_2(j)
	\label{heu}
\end{equation}
with $\alpha \in [0,2]$. In the following, illustrating the proposal, we will restrict considerations to the simplest case with $\alpha=1$. \\
It is worth to comment how the introduction of heuristics as above may be regarded as a `mean field like' procedure, where only pairwise, nearest neighbour interactions are considered (which is the case, for example, in Ising models). The quantity $H$ itself can be interpreted as a \textit{potential}, once changed in sign. One may notice that procedures based on optimization of \textit{Hamiltonians} have already been thoroughly applied to the clustering problem (e.g. \cite{pott.mod04} \cite{influ.liu10}). Indeed, with a terminology drawing on this parallel, a key difference in our approach is that we are defining and using \textit{local} potentials, whereas the traditional approach involves the optimization of a \textit{global} potential.
\subsection{Further comments on the heuristics}
\label{comments}
Given that the heuristics, in the form introduced so far, were only intuitively justified to be reliable metrics for our aim, it is plenty of possible modifications, simplifying or generalizing the particular version given in (\ref{heu1}) and (\ref{heu2}). 

We will take in consideration a few cases which may be interesting for some particular applications. As the first, whenever a speed-up in the computation of the heuristics is required, it is envisaged the possibility to slightly change the definition of $com$ (\ref{comd}), so to skip operations on \textit{positive} (or, equivalently, \textit{negative}) terms. Therefore, this version of the algorithm could use e.g.:
\begin{equation} com^+(i,j)=
	\begin{cases} 
	 0 \;\;\;\;\;\; & \mbox{(if $i$ and $j$ belong to different communities )} \\
	 +1 \;\;\;\;\;\; & \mbox{(if $i=j$ $\vee$ if $i$ and $j$ belong to same cluster)}
	\end{cases}
	\label{comd1}
\end{equation}
or viceversa for $com^-(i,j)$. Steps involving null terms in the computation of $H_1$ and $H_2$ would be excluded by conditional restraints. 

A more interesting case is given by \textit{directed} graphs (i.e. `digraphs'). In fact, to keep the general case as simple as possible, we have always avoided directionality considerations in (\ref{heu1}) and (\ref{heu2}), by using the averaged term $(a_{ij}+a_{ji})/2$. Intuitively, this is equivalent to the replacement of multiple directed (weighted) edges, for each couple of nodes, with a single undirected weighted edge. Even if approaches like this have been applied to highly successful analyses of naturally directed graphs \cite{al-ba99}, it is well recognized how intrinsic directional features may add insight to static \cite{dirweb.brod00} or dynamic \cite{krap01} analyses of networks. Notice that the heuristics introduced may be readily generalized to include different expressions for an `in-metric' $H^{in}_{1,2}$ as well as an `out-metric' $H^{out}_{1,2}$. I.e. for the inner case:
\begin{align}
 H_1^{in}(j) &= \frac{1}{d^{in}_T(j)} \sum_{{i}} {a}_{ij} \; {com(i,j)}
 \label{hin1} \\
 H_2^{in}(j) &= \frac{1}{[d^{in}_T(j)]^2} \sum_{i \neq j} a_{ij} \;  com(i,j) Q \; d^{in}_T(i) H^{in}_1(i) 
\label{hin2}
\end{align}
Now, considerations about robustness of products of inner and outer quantities, for clustering procedures, may apply to this case. In fact, multiplying $H_1^{in}$ and $H_1^{out}$, the product ($H_1'$) closely resembles\footnote{The similarity of these quantities does not imply similarity in their usage, as in \cite{gdl.zhang12} the vertex-cluster affinity (and its derivatives) are directly used for the agglomerative step of the algorithm, whereas we use them only to classify the quality of single-node attributions to clusters.} the \textit{vertex-cluster affinity}, employed in \cite{gdl.zhang12} for the \textit{graph degree-linkage} method, where the cluster would here be the neighborhood $\mathcal{N}$ of each critical vertex. The contribution from $H_2'$ can instead be seen as an improvement of this affinity. Therefore, drawing on these previous results, we claim that a robust implementation of our procedure in directional cases uses the node heuristics:
\begin{align}
 H_1'(j) &:= H_1^{in}H_1^{out} \\
 H_2'(j) &:= H_2^{in}H_2^{out}
\label{hdir}
\end{align}
and the obvious generalization of (\ref{heu}) for their combination.

It is still left open, the possibility to drastically change the form of the heuristic metrics adopted. For example, given that $H_1$ is claimed to be a measure of the membership degree of node $j$ to its initial community, one could recall how this indication is embedded in the elements of the \textit{membership matrix}, as defined in \cite{pattern81}. However, the additional definitions of `positions' and `distances' in a metric space, required in the definition of this matrix, may be rather artificial for some graphs \cite{fortu10}.\\
Again modifying preliminary definitions: $Q$, as given in (\ref{qd}), may be considered a rough figure of merit for the degree ratio in (\ref{terms2}). E.g. one could assume a Gaussian behaviour in the degree distribution, and thus suppose $Q$ to be in the form of a standard deviation\footnote{Notice that the expected value for the population is trivially $<\delta_D>=0$.}:
\begin{equation} 
   Q^2= \frac{2}{n(n-1)} \sum_{\substack{i,j \\ i\neq j}} [\delta_D(i,j)]^2,
	\label{qd2}
\end{equation}
	\begin{equation}
	\mbox{with:   } \delta_D(i,j)=(d_T(i)-d_T(j))
	\label{deltad}
\end{equation}
and therefore it may be objected that: 
\begin{equation} 
   M' := exp(-\delta_D(i,j)^2/Q^2)
	\label{ratiod2}
\end{equation}
is a more reliable measure as a \textit{degree distance} among nodes $i$ and $j$ and should replace $M$ (\ref{terms2}) in (\ref{heu2}). Notice that (\ref{ratiod2}) provides a measure, on the strength of the connection between the nodes, resembling the \textit{dimensionality reduction} procedure invoked for graph construction in \cite{eigmaps03} and related works. However, two considerations hold. The first and more important is that computing the quantity in (\ref{qd2}) is a computational problem much more costly (and in some cases even tricky \cite{smpvar83}), than the linear scan required for computing (\ref{qd}). A second noticeable problem is that the naive introduction of this `standard variance' form for $Q$ does not fit well our requirements. Indeed, it is introduced an unwanted symmetry: $M'$ is a factor reducing the importance of $H_2$, indifferently of whose node is the degree centrality increasing (whereas in $M$ this was true only in the situation $d_T(j)>>d_T(i)$). In formulas, where $\epsilon$ is $O(e^{-n^2})$:
\begin{align} 
   lim_{d_T(j)/d_T(i)\rightarrow \infty} M &\stackrel{}{\longrightarrow}0 \\
   lim_{|d_T(j)-d_T(i)|\rightarrow \infty} M' &\stackrel{}{\longrightarrow} \epsilon
	\label{lims.M}
\end{align}
There are certainly various possibilities to solve the issue: e.g. introducing further parameters in (\ref{deltad}), or defining it differently. However, in our opinion this unnecessarily complicates the global picture, and therefore move on to test numerically the performance of the heuristics outlined.\\

\begin{table}
	\centering
		\begin{tabular}{c | c c c c c | c}
			\hline \hline  
			\\[-2.5pt]
& $H_1(i)$ & $d_T(i)$ & $com(i,j)$ & $Q$ & $... \sum_{i} {\left( {a}_{ij}+{a}_{ji} \right) ... }$ & \textit{Total} 
  \\ [0.5pt]
			\hline
			\\[-2.5pt]
$H_1(i)$ & - & $\mathcal{O}(m/n)$ & c.g.? & - & $\mathcal{O} (m/n)$ & $\mathcal{O} (m/n)$ \\ 
$H_2(i)$ & $\mathcal{O}(m/n)$ & $\mathcal{O}(m/n)$ & c.g.? & $\mathcal{O} (1)$ & $\mathcal{O} (m/n)$ & $\mathcal{O} (\frac{m}{n}+1)$ \\ \hline \\[-2.5ex]
size & $\mathcal{O}(n)$ & $\mathcal{O}(n)$ & $\mathcal{O}(n)$ & $\mathcal{O}(1)$ & $\mathcal{O} (m)$ & $\mathcal{O} (m+n)$\\ [0.5ex]
\hline 
		\end{tabular}
		\smallskip
		\caption{Analysis of computational complexity (average, per node) and memory usage (globally) of the quantities involved in the calculation of the heuristics in Eqs.(\ref{heu1}) and (\ref{heu2}), for an undirected graph. \textit{c.g.?} indicates that the complexity of this step depends on the coarse algorithm applied. $com(i,j)$ is supposed to be retrieved from a stored vector of single-node assignments.}
	\label{tab:complex}
\end{table}
\subsection{Discussion about implementation}
\label{implement}

We are now left with checking the respondency of our proposal for a heuristic, to the requirements stated in Pts. 2-3 of Prop.\ref{assmp}.
Observing Table \ref{tab:complex}, it is easy to see that $H_1$ has a complexity of $\mathcal{O}(m)$ and $H_2$ has a complexity of $\mathcal{O}(m+n)$, under the following assumptions: the graph is undirected and stored as an ordered \textit{edgelist}\footnote{If not, an additional step with complexity $\mathcal{O}(m \; log \; m)$ must be taken into account}, coarse communities have already been calculated and stored in a vector. Notice how redundant terms in the two heuristics can ease the subsequent calculation of both quantities $H_{1,2}$. Such a complexity is a reasonably good result: one of the fastest coarse algorithms for community detection runs with complexity $\mathcal{O}(n + m)$ on sparse graphs. Additionally, operations leading to the heuristics' complexity are very basic, thus we envisage very low factors. 

In order to perform a test for the multi-step scheme, following also Fig.\ref{fig:scheme}, two elements are required to be explicitated.\\
A \textsl{coarse grain algorithm} for the first step. We chose to use the \textit{fast Newman} (FN) approach \cite{fast.newman04} with a greedy modularity optimization of the \textit{modularity}, as suggested in \cite{mod.cnm04}. Within this implementation, it is known to run in $\mathcal O (n \; log^2n)$ on sparse graphs. This method is of widespread adoption in the literature\footnote{Its combined simplicity and robustness make the FN method still very popular, even if several works have started to point out its ineffectiveness for specific cases \cite{critiq.good10}, \cite{badq.kehagias12}.} and in several network analysis softwares. \\
A \textsl{refinement algorithm} for the final step. In this case we introduced a modified \textit{Girvan-Newman} (GN) method, based on the \textit{edge betweenness}: a perfect example of an algorithm unfeasible to be used straightforward for large networks, as it requires $\mathcal O (n^3)$ time (sparse case). The original version of this algorithm was not intended to perform single node re-assignments \cite{girvan.newman02}, so that it is here modified, even though keeping the same local measure as the working principle. In brief, here the edge betweenness is calculated only for \textit{critical edges}, i.e. those edges linking couples of nodes, of whose at least one is \textit{critical}. The last edge to be removed, before a node is isolated, is also the one ruling the community assignment\footnote{Specifically, we progressively remove critical edges with high edge-betweenness. The algorithm has three hierarchical rules to assign node to the refined \textit{community vector}: i) if a critical node $i$ is left with one only edge linking it to a non-critical node $j$ (i.e. $a_{ij}+a_{ji} \neq 0$), $i$ acquires the same community assignment of $j$: $i\in \mathcal{C}_k$ \textsl{iff} $j\in \mathcal{C}_k$; ii) for critical nodes pointing to each other, before becoming isolated by the edge-removal procedure (`queued nodes'), it is attempted the creation of a new community; iii) if this attempt fails, the \textit{transitivity principle} introduced in the text is used to infer the non-critical node ruling the community assignment. The full algorithm in pseudocode can be found in the Appendix.}.  
Notice that the refinement algorithm used is allowed both to eventually shrink the number of clusters composing the final partitioning, \textsl{and} to create new clusters, eventually not resolved by the coarse step. \\
The adoption of a refinement step poses a non-trivial problem: given unawareness of the percentage of nodes classified in the wrong cluster by the coarse algorithm, how many nodes must be `refined'analyzed, among those scoring worse in $H$? That is, we need to impose a threshold to the heuristics (see Fig.\ref{fig:scheme}), selecting as critical nodes only those having a lower value of $H$. In our opinion, this point requires a good insight about the structure of the network, and if investigated, can provide interesting results. A pragmatic and prudential solution is: pose the threshold in $H$ as high, as the additional computational time, required for refinement, is considered feasible by the adopter. For numerical tests below, we will adopt instead an `absolute' approach: the refinement algorithm will be run on all, and only, those nodes having negative values of $H$.
\begin{figure}
	\centering
		\includegraphics[width=9cm, height=5.5cm]{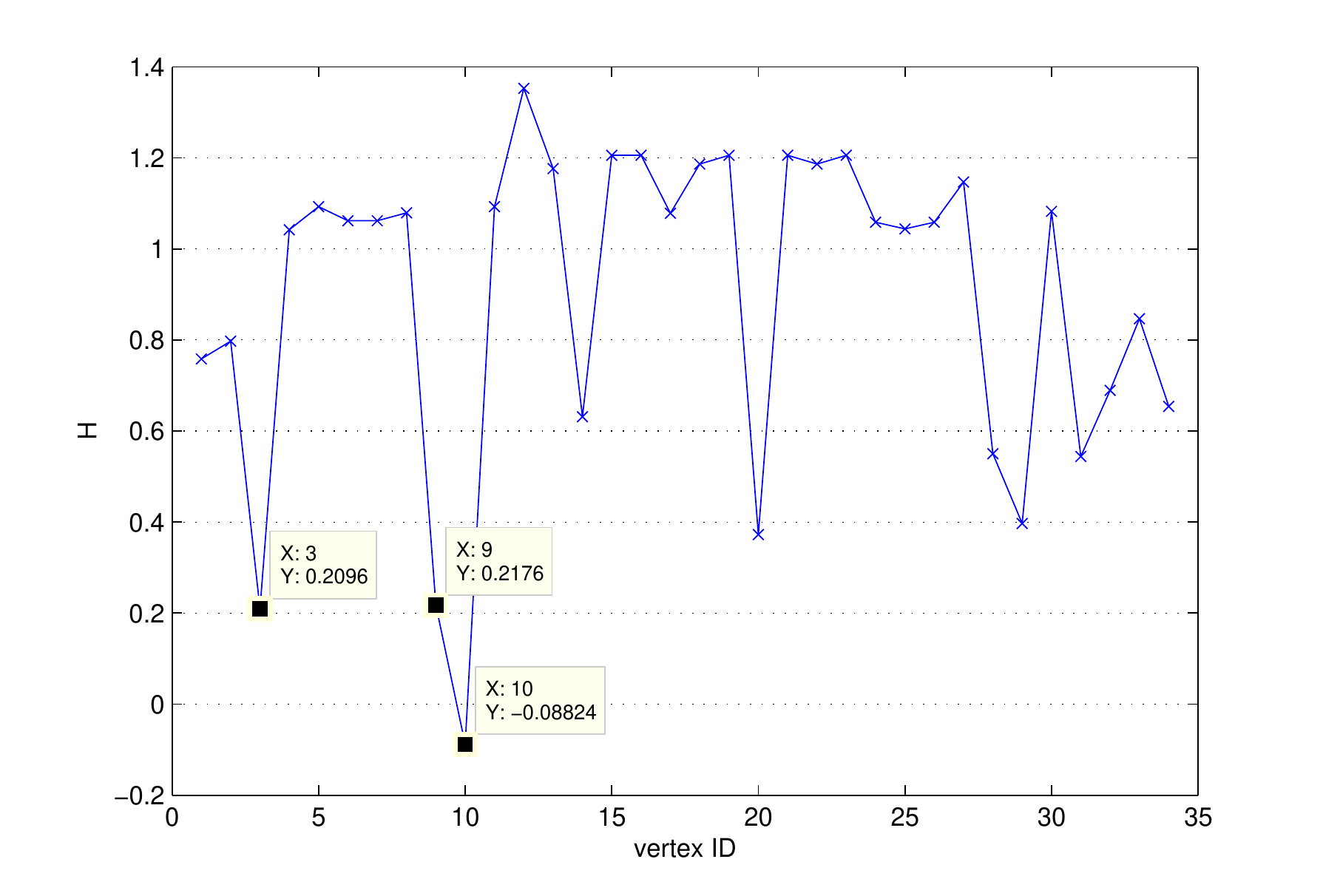}
	\caption{Combined heuristic $H$ (Eq.\ref{heu}) for the Zachary's karate club case. Datatips are displayed in the first case for those vertices exhibiting the lower scoring, and therefore the most critical ones. Vertices IDs are the same as in \cite{girvan.newman02}.}
	\label{fig:hqk}
\end{figure}
\section{Preliminary tests}
\label{test}
This paragraph is devoted to show how the particular implementation of a multi-step scheme (as outlined in Par.\ref{implement}) works for a real case, and in particular to test if the heuristics, introduced so far, are capable of satisfying the requirements stated at the beginning of this section. 

\begin{figure}
	\centering
		\includegraphics[width=11cm, height=5cm]{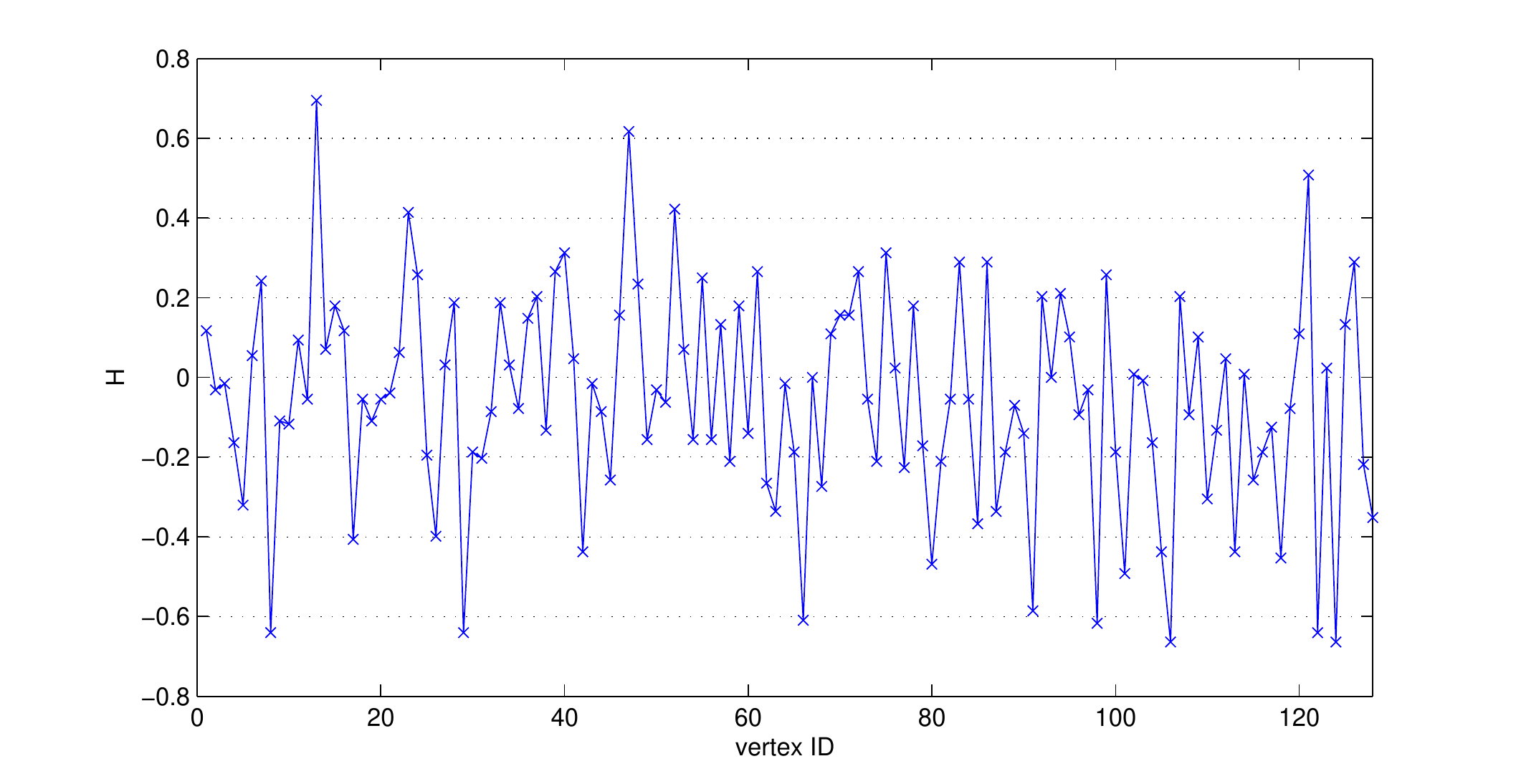}
	\caption{Combined heuristic $H$ for the \textit{Girvan-Newman benchmark} cited in the text. Generated using the code as in \cite{lanci.bench08}, with average degree $\bar{d}=16$, and mixing parameter $\mu=0.6$.}
	\label{fig:hgn}
\end{figure}

\textit{Test-cases}. We have chosen to focus on the split of a \textit{karate club} in two different `communities', studied in \cite{karate.zach77}. This example fits well a preliminary, qualitative discussion, because it is small enough ($n$=34) to let us follow in detail the performance of the heuristics, yet it is complex enough to pose difficulties for the fast coarse algorithm chosen \cite{mod.newman06}.
In Fig.\ref{fig:hqk} we plot the sum of the heuristics $H_1$ and $H_2$ for all the vertices of the karate club network, after a coarse assignment of clusters has been performed through the application of the FN algorithm. It is immediately evident how almost all of the nodes have positive values of both $H_{1,2}$. This confirms that $H$ captures the good performance of the FN algorithm in this test. We can also state that our core claim is satisfied: the node $\#10$, known to be misclassified by the coarse algorithm \cite{fast.newman04}, is the one scoring worse, and even has a negative $H$, as shown in Fig.\ref{fig:hqk}. Notice also that the GN refinement correctly classifies this node, displacing it into the `right' cluster. Recalling (\ref{part.dist}), the coarse method has in this case a distance of $D\cong 0.029$ from the partition found by our scheme: this distance can be understood as the \textit{improvement} provided for the solution. Noticeably, for this specific case, the GN method is known to classify incorrectly node $\#3$ \cite{girvan.newman02}, which actually ranked worst, immediately after node $\#10$, in the $H$ scoring. This further suggests how the heuristic proposed is indeed efficient, in sorting nodes with uncertain cluster assignment.\\
As a counter-example, in Fig.\ref{fig:hgn} we report the heuristics, calculated after the same coarse step, run on an artificial network\footnote{Created through the benchmark package available at: \url{goo.gl/Btp70b}} with $n$=128. This network resembles a `Girvan-Newman benchmark' with communities of variable size, and with parameters which are known to make the community assignment fail, when performed by the FN algorithm \cite{danon06}\cite{lanci.bench08}. It is immediately evident how the average scoring of the heuristic $H$ is much worse than the previous case, and how most of the nodes exhibit negative scoring. This indicates again that the heuristics scouts nodes misclassified in the coarse step. 
\section{Conclusions}
Summarizing the main results of this work: we have proposed the adoption of a multi-step scheme, to improve the results of clustering algorithms, with a particular focus on community detection. This scheme basically includes: the adoption of a (state-of-art) fast, coarse algorithm for the first step; an accurate refinement algorithm, specifically adapted for this purpose; to bridge these two elements, a novel set of heuristic metrics. These last ones are the core of the proposal: they are intended to scout those nodes potentially tricky in the cluster assignment, and thus worth to be analyzed by the refinement step. We have shown, with the aid of test-cases, that the heuristic introduced satisfies the requirements of being computable with low time-complexity, and may efficiently retrieve those nodes which turn out to `deceive' the less accurate algorithms.\\
In future developments there is the plan to systematically investigate to what extent our approach reveals useful for application to real world and computer generated networks (thus identifying its limits). In particular, the aim will be about large scale networks, for which it may also be unknown the `true partitioning' (whether obtained via a direct observation, or as provided by the application of the refinement to the whole network). In this case the only possible check would be the comparison with results, as provided by different fast algorithms. Another direction, for further analyses, is given by the limitations already found for modularity-based approaches \cite{critiq.good10}: we claim that our multi-step strategy may (partially) solve the degeneracies displayed by these approaches for particular cases. Verification of this conjecture could lead to important applications.
\appendix

\section*{Appendix A}
It is reported below the pseudocode related to the \textit{GN refinement} used for numeric experiments in the text (Par.\ref{test}). 

\lstset{
escapeinside={*}{*}, 
basicstyle=\scriptsize\ttfamily,
numbers=none, 
numbersep=2pt,
numberstyle=\tiny,
language=C,
morekeywords={assign, remove, find, extract, move, sort}
}

\begin{lstlisting}
*program `GN refinement' (coarse \textit{com-vector})*{
  for (each *\em node* in *\em graph*){
    if (*H* (*\em node*) > *\em threshold-1*){
      assign (*\em node* to *\em critical-nodes*)
      assign (edges connecting the node to *\em critical-edges*)}}
  do edge_betwenness (*\em critical-edges*)
  sort (*\em critical-edges*, descending edge_betweenness)
  while (there are *\em critical-edges* left){
    *\em newgraph* = remove (*\em critical-edge* with highest betweenness from *\em graph*)
    if (*\em critical-edge* connects two nodes, of which one *$\notin$ \em critical-nodes*){
      *\em header-node*[*\em critical-node*]= its neighbour-node from removed edge
      assign increasing *\em priority*[*\em critical-node*]}
    find (*\em critical-nodes* left with one only edge)
    for (each of these *\em nodes*){
      if (the only *\em neighbour-node* left *$\in $* *\em critical-nodes*){
        assign (*\em node* to *\em queued-nodes*)
        *\em follower-node*[*\em neighbour-node*]=*\em node*}
      else{
        *\em{com-vector}*[*\em node*]=*\em{com-vector}*[*\em neighbour-node*]
        assign(*\em node* to *\em solved-nodes*)}}}
  *\em listed-nodes* = merge (*\em queued-nodes* with corresponding *\em neighbour-nodes*)
  extract (from *\em graph* the *\em subgraph* including all and only the *\em listed-nodes*)
  *\em comps* = connected_components (*\em subgraph*)
  for (each *\em comp*){
    if (sizeof *\em comp* > *\em threshold-2*){
      assign(nodes in *\em comp* to novel community)
      remove(queued nodes in *\em comp* from *\em queued-nodes*)}}
  for (each of the remaining *\em queued-nodes*){
    if (*\em neighbour-node*[*\em queued-node*] *$\in $* *\em solved-nodes*){
      *\em{com-vector}*[*\em queued-node*]=*\em{com-vector}*[*\em neighbour-node*[*\em queued-node*]]
      assign (*\em queued-node* to *\em solved-nodes*) & remove(*\em node* from *\em queued-nodes*)}}
    while (there are *\em queued-nodes* left){
      move to next(*\em header-node* with highest *\em priority*, connecting one *\em queued-node*)
      *\em{com-vector}*[corresponding *\em queued-node*]=*\em{com-vector}*[its *\em header-node*]
      assign (*\em queued-node* to *\em solved-nodes*) & remove(*\em node* from *\em queued-nodes*)
      for (all *\em follower-nodes*[this *\em solved-node*]){
        *\em{com-vector}*[*\em follower-node*]=*\em{com-vector}*[*\em solved-node*]
        assign (*\em follower-node* to *\em solved-nodes*) & remove(*\em node* from *\em queued-nodes*)}}
  return(new *\textit{com-vector}*)}
\end{lstlisting}

%
\bibliography{HeuAlg}

\begin{thebibliography}{10}

\bibitem{net.brandes05}
U.~Brandes and T.~Erlebach, {\em Network analysis: methodological foundations},
  vol.~3418.
\newblock Springer, 2005.

\bibitem{fortu10}
S.~Fortunato, ``Community detection in graphs,'' {\em Physics Reports},
  vol.~486, no.~3, pp.~75--174, 2010.

\bibitem{info.fortu04}
S.~Fortunato, V.~Latora, and M.~Marchiori, ``Method to find community
  structures based on information centrality,'' {\em Physical Review E},
  vol.~70, no.~5, p.~056104, 2004.

\bibitem{intro.newman11}
M.~Newman, ``Communities, modules and large-scale structure in networks,'' {\em
  Nature Physics}, vol.~8, no.~1, pp.~25--31, 2011.

\bibitem{bio.jeong00}
H.~Jeong, B.~Tombor, R.~Albert, Z.~N. Oltvai, and A.-L. Barab{\'a}si, ``The
  large-scale organization of metabolic networks,'' {\em Nature}, vol.~407,
  no.~6804, pp.~651--654, 2000.

\bibitem{web.broder97}
A.~Z. Broder, S.~C. Glassman, M.~S. Manasse, and G.~Zweig, ``Syntactic
  clustering of the web,'' {\em Computer Networks and ISDN Systems}, vol.~29,
  no.~8, pp.~1157--1166, 1997.

\bibitem{info.rosvall07}
M.~Rosvall and C.~T. Bergstrom, ``An information-theoretic framework for
  resolving community structure in complex networks,'' {\em Proceedings of the
  National Academy of Sciences}, vol.~104, no.~18, pp.~7327--7331, 2007.

\bibitem{mod.newman06}
M.~E. Newman, ``Modularity and community structure in networks,'' {\em
  Proceedings of the National Academy of Sciences}, vol.~103, no.~23,
  pp.~8577--8582, 2006.

\bibitem{eiratio.krack88}
D.~Krackhardt and R.~N. Stern, ``Informal networks and organizational crises:
  An experimental simulation,'' {\em Social psychology quarterly},
  pp.~123--140, 1988.

\bibitem{multires.ronhovde09}
P.~Ronhovde and Z.~Nussinov, ``Multiresolution community detection for
  megascale networks by information-based replica correlations,'' {\em Physical
  Review E}, vol.~80, no.~1, p.~016109, 2009.

\bibitem{mod.cnm04}
A.~Clauset, M.~E. Newman, and C.~Moore, ``Finding community structure in very
  large networks,'' {\em Physical review E}, vol.~70, no.~6, p.~066111, 2004.

\bibitem{fast.blondel08}
V.~D. Blondel, J.-L. Guillaume, R.~Lambiotte, and E.~Lefebvre, ``Fast unfolding
  of communities in large networks,'' {\em Journal of Statistical Mechanics:
  Theory and Experiment}, vol.~2008, no.~10, p.~P10008, 2008.

\bibitem{critiq.good10}
B.~H. Good, Y.-A. de~Montjoye, and A.~Clauset, ``Performance of modularity
  maximization in practical contexts,'' {\em Physical Review E}, vol.~81,
  no.~4, p.~046106, 2010.

\bibitem{ker.lin70}
B.~Kernighan and S.~Lin, ``An efficient heuristic procedure for partitioning
  graphs,'' {\em Bell system technical journal}, 1970.

\bibitem{eigen.newman06}
M.~E. Newman, ``Finding community structure in networks using the eigenvectors
  of matrices,'' {\em Physical Review E}, vol.~74, no.~3, p.~036104, 2006.

\bibitem{graphintro.west}
D.~B. West {\em et~al.}, {\em Introduction to graph theory}, vol.~2.
\newblock Prentice hall Englewood Cliffs, 2001.

\bibitem{digraph.malliaros13}
F.~D. Malliaros and M.~Vazirgiannis, ``Clustering and community detection in
  directed networks: A survey,'' {\em Physics Reports}, vol.~533, no.~4,
  pp.~95--142, 2013.

\bibitem{mmbdft14}
A.~A. Gentile, A.~Corallo, C.~Bisconti, and L.~Fortunato, ``Proposal for
  heuristics-based refinement in clustering problems,'' in {\em Proceedings of
  the SOCNET '14 Workshop (\texttt{to be published})}, University of Bamberg,
  2014.

\bibitem{clust.kann04}
R.~Kannan, S.~Vempala, and A.~Vetta, ``On clusterings: Good, bad and
  spectral,'' {\em Journal of the ACM (JACM)}, vol.~51, no.~3, pp.~497--515,
  2004.

\bibitem{clust.qual.kri}
H.-P. Kriegel and M.~Pfeifle, ``Measuring the quality of approximated
  clusterings,'' in {\em BTW}, vol.~5, pp.~415--424, 2005.

\bibitem{part.qual.rob}
C.~Robardet, F.~Feschet, and N.~Nicoloyannis, ``An experimental study of
  partition quality indices in clustering,'' in {\em Proceedings of the 4th
  European Conference on Principles of Data Mining and Knowledge Discovery},
  PKDD '00, (London, UK), pp.~599--604, Springer-Verlag, 2000.

\bibitem{sna.wasser94}
S.~Wasserman, {\em Social network analysis: Methods and applications}, vol.~8.
\newblock Cambridge university press, 1994.

\bibitem{pott.mod04}
J.~Reichardt and S.~Bornholdt, ``Detecting fuzzy community structures in
  complex networks with a potts model,'' {\em Physical Review Letters},
  vol.~93, no.~21, p.~218701, 2004.

\bibitem{influ.liu10}
S.~Liu, L.~Ying, and S.~Shakkottai, ``Influence maximization in social
  networks: An ising-model-based approach,'' in {\em Communication, Control,
  and Computing (Allerton), 2010 48th Annual Allerton Conference on},
  pp.~570--576, IEEE, 2010.

\bibitem{al-ba99}
A.-L. Barab{\'a}si and R.~Albert, ``Emergence of scaling in random networks,''
  {\em Science}, vol.~286, no.~5439, pp.~509--512, 1999.

\bibitem{dirweb.brod00}
A.~Broder, R.~Kumar, F.~Maghoul, P.~Raghavan, S.~Rajagopalan, R.~Stata,
  A.~Tomkins, and J.~Wiener, ``Graph structure in the web,'' {\em Computer
  networks}, vol.~33, no.~1, pp.~309--320, 2000.

\bibitem{krap01}
P.~Krapivsky, G.~Rodgers, and S.~Redner, ``Degree distributions of growing
  networks,'' {\em Physical Review Letters}, vol.~86, no.~23, p.~5401, 2001.

\bibitem{gdl.zhang12}
W.~Zhang, X.~Wang, D.~Zhao, and X.~Tang, ``Graph degree linkage: agglomerative
  clustering on a directed graph,'' in {\em Computer Vision--ECCV 2012},
  pp.~428--441, Springer, 2012.

\bibitem{pattern81}
J.~C. Bezdek, {\em Pattern recognition with fuzzy objective function
  algorithms}.
\newblock Kluwer Academic Publishers, 1981.

\bibitem{eigmaps03}
M.~Belkin and P.~Niyogi, ``Laplacian eigenmaps for dimensionality reduction and
  data representation,'' {\em Neural computation}, vol.~15, no.~6,
  pp.~1373--1396, 2003.

\bibitem{smpvar83}
T.~F. Chan, G.~H. Golub, and R.~J. LeVeque, ``Algorithms for computing the
  sample variance: Analysis and recommendations,'' {\em The American
  Statistician}, vol.~37, no.~3, pp.~242--247, 1983.

\bibitem{fast.newman04}
M.~E. Newman, ``Fast algorithm for detecting community structure in networks,''
  {\em Physical Review E}, vol.~69, no.~6, p.~066133, 2004.

\bibitem{badq.kehagias12}
A.~Kehagias, ``Bad communities with high modularity,'' {\em preprint
  arXiv:1209.2678}, 2012.

\bibitem{girvan.newman02}
M.~Girvan and M.~E. Newman, ``Community structure in social and biological
  networks,'' {\em Proceedings of the National Academy of Sciences}, vol.~99,
  no.~12, pp.~7821--7826, 2002.

\bibitem{karate.zach77}
W.~W. Zachary, ``An information flow model for conflict and fission in small
  groups,'' {\em Journal of anthropological research}, pp.~452--473, 1977.

\bibitem{danon06}
L.~Danon, A.~D{\'\i}az-Guilera, and A.~Arenas, ``The effect of size
  heterogeneity on community identification in complex networks,'' {\em Journal
  of Statistical Mechanics: Theory and Experiment}, vol.~2006, no.~11,
  p.~P11010, 2006.

\bibitem{lanci.bench08}
A.~Lancichinetti, S.~Fortunato, and F.~Radicchi, ``Benchmark graphs for testing
  community detection algorithms,'' {\em Physical Review E}, vol.~78, no.~4,
  p.~046110, 2008.

\end{thebibliography}
\bibliographystyle{ieeetr}
%
%
%
%
%
%
%
\clearpage
\addtocmark[2]{Author Index} 
\renewcommand{\indexname}{Author Index}
\renewcommand{\indexname}{Subject Index}
\end{document}